\documentclass[twocolumn,showpacs,preprintnumbers,amsmath,amssymb]{revtex4}

\usepackage{graphicx}
\usepackage{dcolumn}
\usepackage{bm}

\begin{document}
\title{Generalized BBV Models for Weighted Complex Networks}
\author{Bo Hu$^{1}$}
\email{hubo25@mail.ustc.edu.cn}
\author{Gang Yan$^{2}$}
\author{Wen-Xu Wang$^{1}$}
\author{Wen Chen$^{1}$}
\affiliation{%
$^{1}$Nonlinear Science Center and Department of Modern Physics,
University of Science and Technology of China, Hefei, 230026, PR
China  \\
$^{2}$Department of Electronic Science and Technology, University
of Science and Technology of China, Hefei, 230026, PR China
}%
\date{\today}

\begin{abstract}
We will introduce two evolving models that characterize weighted
complex networks. Though the microscopic dynamics are different,
these models are found to bear a similar mathematical framework,
and hence exhibit some common behaviors, for example, the
power-law distributions and evolution of degree, weight and
strength. We also study the nontrivial clustering coefficient $C$
and tunable degree assortativity coefficient $r$, depending on
specific parameters. Most results are supported by present
empirical evidences, and may provide us with a better description
of the hierarchies and organizational architecture of weighted
networks. Our models have been inspired by the weighted network
model proposed by Alain Barrat \emph{et al.} (BBV for short), and
can be considered as a meaningful development of their original
work.
\end{abstract}

\maketitle
\section{Introduction}
The recent few years have witnessed a great development in physics
community to explore and characterize the underlying laws of
complex networks, including as diverse as the Internet
\cite{Internet}, the World-Wide Web \cite{WWW}, the scientific
collaboration networks (SCN) \cite{CN1,CN2}, and world-wide
airport networks (WAN)\cite{air1,air2}. Many empirical
measurements have uncovered some general scale-free properties of
those real systems, which motivated a wealth of theoretical
efforts devoted to the characterization and modelling of them.
Since Barab\'asi and Albert introduced their seminal BA model,
most efforts have been contributed to study the network
topological properties \cite{BA}. However, networks as well known
are far from boolean structures, and the purely topological
representation of them will miss important attributes often
encountered in real world. For instance, the amount of traffic
characterizing the connections of communication systems or large
transport infrastructure is fundamental for a full description of
them \cite{top10}. More recent years, the availability of more
complete empirical data and higher computation ability permit
scientists to consider the variation of the connection strengths
that indeed contain the physical features of many real graphs.
Weighted networks can be described by a weighted adjacency
$N\times N$ matrix, with element $w_{ij}$ denoting the weight on
the edge connecting vertices $i$ and $j$. As a note, this paper
will only consider undirected graphs where weights are symmetric.
As confirmed by measurements, complex networks often exhibit a
scale-free degree distribution $P(k)\thicksim k^{-\gamma}$ with
2$\leq\gamma\leq$3 \cite{air1,air2}. Interestingly, the weight
distribution $P(w)$ is also found to be heavy tailed in some real
systems \cite{ref1}. As a generalization of connectivity $k_i$,
the vertex strength is defined as
$s_{i}=\sum_{j\in\Gamma(i)}w_{ij}$, where $\Gamma(i)$ denotes the
set of $i$'s neighbors. This quantity is a natural measure of the
importance or centrality of a vertex in the network. For instance,
the strength in WAN provides the actual traffic going through a
vertex and is obvious measure of the size and importance of each
airport. For the SCN, the strength is a measure of scientific
productivity since it is equal to the total number of publications
of any given scientist. Empirical evidence indicates that in most
cases the strength distribution has a fat tail \cite{air2},
similar to that of degree distribution. Highly correlated with the
degree, the strength usually displays scale-free property
$s\thicksim k^{\beta}$ with $\beta\geq1$ \cite{traffic-driven,
empirical}. Driven by new empirical findings, Alain Barrat
\emph{et al.} have presented a simple model (BBV for short) that
integrates the topology and weight dynamical evolution to study
the dynamical evolution of weighted networks \cite{BBV}. An
obvious virtue of their model is its general simplicity in both
mechanisms and mathematics. Thus it can be used as a starting
point for further generalizations. It successfully yields
scale-free properties of the degree, weight and strength, just
depending on one parameter $\delta_{BBV}$ that controls the local
dynamics between topology and weights. Inspired by BBV's work, a
class of evolving models will be presented in this paper to
describe and study specific weighted networks. This paper is
organized as follows: In Section II, we will introduce a
traffic-driven model to mimic the weighted technological networks.
Analytical calculations are in consistent with numerical results.
In Section III, a neighbor-connected model is proposed to study
social networks of collaboration, with the comparison of
simulations and theoretical prediction as well. At the end of each
section, we discuss the differences between the BBV model and
ours, from the microscopic mechanisms to observed macroscopic
properties. We conclude our paper by a brief review and outlook in
Section IV.

\section{Model A}

\subsection{The Traffic-Driven Model for Technological Networks}
The network provides the substrate on which numerous dynamical
processes occur. Technology networks provide large empirical
database that simultaneously captures the topology and the traffic
dynamics taking place on it. We argue that traffic and its
dynamics is a key role for the understanding of technological
networks. For Internet, the information flow between routers
(nodes) can be represented by the corresponding edge weight. The
total (incoming and outgoing) information that each router deals
with can be denoted by the node strength, which also represents
the importance or load of given router. Our traffic-driven model
starts from an initial configuration of $N_0$ vertices fully
connected by links with assigned weight $w_0=1$. The model is
defined on two coupled mechanisms: the topological growth and the
increasing traffic dynamics:

\emph{(i) Topological Growth.} At each time step, a new vertex is
added with $m$ edges connected to $m$ previously existing vertices
(we hence require $N_0>m$), choosing preferentially nodes with
large strength; i.e. a node $i$ is chosen by the new according to
the strength preferential probability:
\begin{equation}
\Pi_{new\rightarrow i}=\frac{s_i}{\sum_ks_k}.
\end{equation}The weight of each new edge is also fixed to
$w_{0}=1$. This strength preferential mechanism have simple
physical and realistic interpretations in Ref. \cite{BBV, WWX}.

\emph{(ii) Traffic Dynamics.} From the start of the network
growing, traffic in all the sites are supposed to constantly
increase, with probability proportional to the node strength
$s_i/\sum_ks_k$ per step. We assume the growing speed of the
network's total traffic as a discrete constant $W$ (each unit can
be considered as an information packet in the case of Internet).
Then in statistic sense the newly created traffic in site $i$ per
step is
\begin{equation}
\Delta W_i=W\frac{s_i}{\sum_ks_k}.
\end{equation}These newly-added packets will be sent out from $i$ to their separate
destinations. Our model does not care their specific destinations
or delivering paths, but simply suppose that each new packet
preferentially takes the route-way with larger edge weight (data
bandwidth of links), i.e. with the probability $w_{ij}/s_i$, and
it hence will increase the traffic (strength) in the corresponding
neighbor $j$ of node $i$. It is a plausible mechanism in many
real-world webs. For instance, in the case of the airport
networks, the potential passenger traffic in larger airports (with
larger strength, often located in important cities) will be
usually greater than that in smaller airports, and busy airlines
often get busier in development. For Internet, routers that have
larger traffic handling capabilities are responsible to deal with
more information packets. Also, the route-ways with broader data
bandwidth will get busier. Admittedly, this ``busy get busier"
scenario is intuitive in physics, though perhaps not strict in
mathematics.

\subsection{Analytical Results vs. Numerical Simulations}
The model time is measured with respect to the number of nodes
added to the graph, i.e. $t=N-N_0$, and the natural time scale of
the model dynamics is the network size $N$. In response to the
demand of increasing traffic, the systems must expand in topology.
With a given size, one technological network assumably has a
certain ability to handle certain traffic load. Therefore, it
could be reasonable to suppose for simplicity that the total
weight on the networks increases synchronously by the natural time
scale. That is why we assume $W$ as a constant. This assumption
also bring us the convenience of analytical discussion \cite{WWX}.
By using the continuous approximation, we can treat $k, w, s$ and
the time $t$ as continuous variables. The time evolution of the
weights $w_{ij}$ can be computed analytically as follows:
\begin{eqnarray}
\frac{dw_{ij}}{dt}&=&\Delta W_i\frac{w_{ij}}{s_i}+\Delta
W_j\frac{w_{ij}}{s_j} \nonumber\\
&=&W\frac{s_i}{\sum_ks_k}\frac{w_{ij}}{s_i}+
W\frac{s_j}{\sum_ks_k}\frac{w_{ij}}{s_j} \nonumber\\
&=&2W\frac{w_{ij}}{\sum_ks_k}.
\end{eqnarray} The term $\Delta W_iw_{ij}/s_i$ represents the
contribution to weight $w_{ij}$ from site $i$. Considering
\begin{equation}
\sum_ks_k\approx(2W+2m)t,
\end{equation}one can rewrite the above evolution equation as:
\begin{equation}
\frac{dw_{ij}}{dt}=\frac{W}{W+m}\frac{w_{ij}}{t}.
\end{equation}
\newcommand{\zwfs}[2]{\frac{\;#1\;}{\;#2\;}}
The link (i,j) is created at $t_{ij}=max(i,j)$ with initial
condition $w_{ij}(t_{ij})=1$, so that
\begin{equation}
w_{ij}(t)=\left(\frac{t}{t_{ij}}\right)^{W/(W+m)}.
\end{equation} Further, we can obtain the evolution equations for $s_i(t)$ and
$k_i(t)$:
\begin{eqnarray}
\frac{ds_i}{dt}&=&\sum_{j\in \Gamma(i)}\frac{dw_{ij}}{dt}+m\frac{s_i}{\sum_ks_k} \nonumber\\
&=&2W\frac{\sum_jw_{ij}}{\sum_ks_k}+\frac{ms_i}{\sum_ks_k} \nonumber\\
&=&(2W+m)\frac{s_i}{\sum_ks_k} \nonumber\\
&=&\frac{2W+m}{2W+2m}\frac{s_i}{t},
\end{eqnarray}and
\begin{equation}
\frac{dk_i}{dt}=m\frac{s_i}{\sum_ks_k}=\frac{ms_i}{2(W+m)t}.
\end{equation}These equations can be readily integrated with
initial conditions $k_i(t_i)=s_i(t_i)=m$, yielding
\begin{eqnarray}
s_i(t)&=&m\left(\zwfs{t}{i}\right)^\frac{2W+m}{2W+2m},\\
k_i(t)&=&m\frac{2W+s_i}{2W+m}.
\end{eqnarray}The strength and degree of vertices are thus related
by the following expression
\begin{equation}
s_i=\frac{2W+m}{m}k_i-2W.
\end{equation}
In order to check the analytical predictions, we performed
numerical simulations of networks created by the present model
with various values of $W$ and minimum degree $m$. In Fig. 1, we
report the average strength $s_i$ of vertices with connectivity
$k_i$ and confirm the validity of Eq. (11).

The time $t_i=i$ when the node $i$ enters the system is uniformly
distributed in $[0,t]$ and the strength probability distribution
can be written as
\begin{equation}
P(s,t)=\frac{1}{t+N_0}\int_{0}^{t}\delta(s-s_i(t))dt_i,
\end{equation}
where $\delta(x)$ is the Dirac delta function. Using equation
$s_i(t)\sim(t/i)^{a}$ obtained from Eq. (9), one obtains in the
infinite size limit $t\rightarrow\infty$ the power-law
distribution $P(s)\sim s^{-\gamma_{_A}}$ (as shown in Fig. 2) with
\begin{equation}
\gamma_{_A}=1+\frac{1}{a}=2+\frac{m}{2W+m}.
\end{equation}Obviously, when $W=0$ the model is topologically
equivalent to the BA network and the value $\gamma_{_A}=3$ is
recovered. For larger values of $W$, the distribution is gradually
broader with $\gamma_{_A}\rightarrow2$ when $W\rightarrow\infty$.
Since $s$ and $k$ are proportional, one can expect the same
behavior of degree distribution $P(k)\sim k^{-\gamma_{_A}}$.
Analogously, the weight distribution can be calculated yielding
the scale-free property $P(w)\sim w^{-\alpha_{_A}}$, with the
exponent $\alpha_{_A}=2+m/W$ (See Fig. 3).

\begin{figure}
\scalebox{0.80}[0.75]{\includegraphics{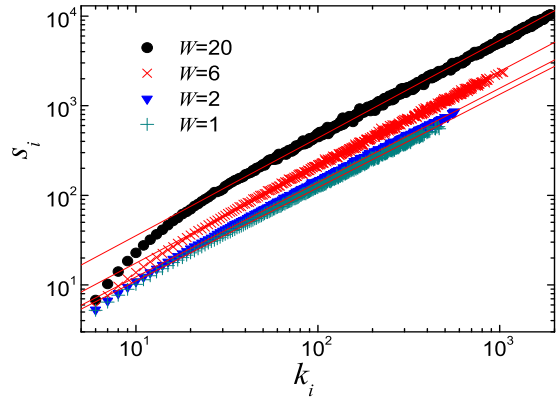}}
\caption{\label{fig:epsart} The average strength $s_i$ of the
nodes with connectivity $k_i$ for different $W$ (log-log scale).
Linear data fittings all give slope $1.00\pm0.02$, demonstrating
the predicted linear correlation between strength and degree.}
\end{figure}

\begin{figure}
\scalebox{0.80}[0.75]{\includegraphics{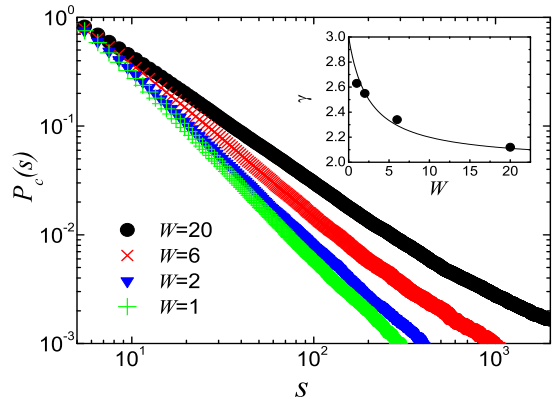}}
\caption{\label{fig:epsart} Cumulative strength probability
distribution $P_c(s)$ with various $W$. Data agree well with the
power-law form $s^{-\gamma_{_A}}$. The inset reports $\gamma_{_A}$
from data fitting (filled circles), in comparison with the
theoretical prediction
$\gamma_{_A}=2+\frac{m}{2W+m}=2+\frac{5}{2W+5}$(line), averaged
over 20 independent networks of size N=7000.}
\end{figure}

\begin{figure}
\scalebox{0.80}[0.75]{\includegraphics{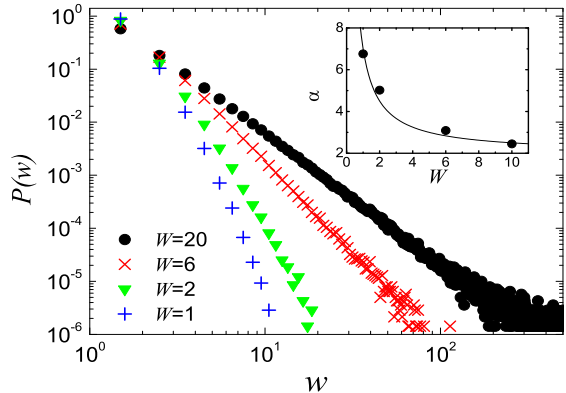}}
\caption{\label{fig:epsart} Weight probability distribution $P(w)$
with various values of $W$. Data are consistent with a power-law
behavior $w^{-\alpha_{_A}}$. In the inset we give the value of
$\alpha_{_A}$ obtained by data fitting (filled circles), together
with the analytical expression $\alpha_{_A}=2+m/W=2+5/W$(line).
The data are averaged over 20 independent realizations of network
size N=7000.}
\end{figure}

\subsection{Clustering and Correlations}
A complete characterization of the network structure must take
into account the level of clustering and degree correlations
present in the network. Information on the local connectedness is
provided by the clustering coefficient $c_i$ defined for any
vertex $i$ as the fraction of connected neighbors of $i$. The
average clustering coefficient $C=N^{-1}\sum_ic_i$ thus expresses
the statistical level of cohesiveness measuring the global density
of interconnected vertices' triples in the network. Further
information can be gathered by inspecting the average clustering
coefficient $C(j)$ restricted to classes of vertices with degree
$k$:
\begin{equation}
C(k)=\frac{1}{NP(k)}\sum_{i/k_i=k}c_i.
\end{equation}
In many networks, $C(k)$ exhibits a power-law decay as a function
of $k$, indicating that low-degree nodes generally belong to well
interconnected communities (high clustering coefficient) while
high-degree sites are linked to many nodes that may belong to
different groups which are not directly connected (small
clustering coefficient). This is generally the signature of a
nontrivial architecture in which hubs (high degree vertices) play
a distinct role in the network. Correlations can be probed by
inspecting the average degree of the nearest neighbors of a vertex
$i$, that is, $k_{nn,i}=k_i^{-1}\sum_{j}k_j$. Averaging this
quantity over nodes with the same degree $k$ leads to a convenient
measure to investigate the behavior of the degree correlation
function
\begin{equation}
k_{nn}(k)=\frac{1}{NP(k)}\sum_{i/k_i=k}k_{nn,i}=\sum_{k'}k'P(k'|k),
\end{equation}
If degrees of neighboring vertices are uncorrelated, $P(k'|k)$ is
only a function of $k'$ and thus $k_{nn}(k)$ is a constant. When
correlations are present, two main classes of possible
correlations have been identified: {\it assortative} behavior if
$k_{nn}(k)$ increases with $k$, which indicates that large degree
vertices are preferentially connected with other large degree
vertices, and {\it disassortative} if $k_{nn}(k)$ decreases with
$k$. The above quantities provide clear signals of a structural
organization of networks in which different degree classes show
different properties in the local connectivity structure. Almost
all the social networks empirically studied show assortative
mixing pattern, while all others including technological and
biological networks are disassortative. The origin of this
difference is not understood yet. To describe these types of
mixing, Newman further proposed some simpler measures, which is
called assortativity coefficients \cite{mixing}. In this paper, we
will also use the formula proposed by Newman \cite{mixing},
\begin{equation}
r=\frac{M^{-1}\sum_ij_ik_i-[M^{-1}\sum_i\frac{1}{2}(j_i+k_i)]^{2}}
{M^{-1}\sum_i\frac{1}{2}(j_i^{2}+k_i^{2})-[M^{-1}\sum_i\frac{1}{2}(j_i+k_i)]^{2}},
\end{equation}
where $j_i$, $k_i$ are the degrees of vertices at the ends of the
$i$th edges, with $i=1,...,M$ ($M$ is the total number of edges in
the observed graph). Simply, $r>0$ means assortative mixing, while
$r<0$ implies disassortative pattern.

\begin{figure}
\scalebox{0.80}[0.75]{\includegraphics{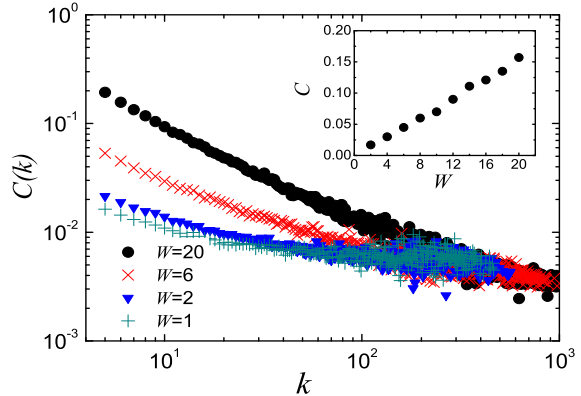}}
\caption{\label{fig:epsart} The scaling of $C(k)$ with $k$ for
various $W$. The data are averaged over 20 independent runs of
network size N=7000. The inset shows the average clustering
coefficient $C$, depending on increasing $W$.}
\end{figure}

\begin{figure}
\scalebox{0.80}[0.75]{\includegraphics{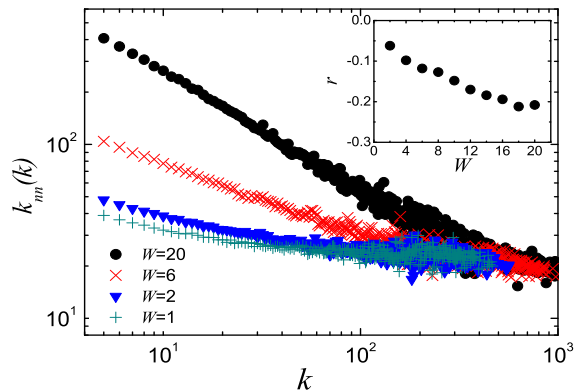}}
\caption{\label{fig:epsart} Average connectivity $k_{nn}(k)$ of
the nearest neighbors of a node depending on its connectivity $k$
for different $W$. The data are averaged over 20 network
realizations of N=7000. The assortativity coefficient $r$
depending on $W$ is shown in the inset.}
\end{figure}

In order to inspect the above properties we perform simulations of
graphs generated by the model for different values of $W$, fixing
$N=7000$ and $m=5$. In the case of clustering, the model also
exhibits properties which are depending on the basic parameter
$W$. More precisely, for small $W$, the clustering coefficient of
the network is small and $C(k)$ is flat. As $W$ increases however,
the global clustering coefficient increases and $C(k)$ becomes a
power-law decay similar to real network data \cite{hierarchy}.
Fig. 4 shows that the increase in clustering is determined by
low-degree vertices. The average clustering coefficient $C$ is
found to be a function of $W$, as shown in its inset. This
numerical result obviously demonstrates the important effect of
traffic on the hierarchical structure of technological networks.
Analogous properties are obtained for the degree correlation
spectrum. For small $W$, the average nearest neighbor degree
$k_{nn}(k)$ is quite flat as in the BA model. The disassortative
character emerges as $W$ increases and gives rise to a power law
behavior of $k_{nn}(k)\sim k^{-\eta}$ (Fig. 5). The assortativity
coefficient $r$ versus $W$, as reported in the inset of Fig. 5,
demonstrates the tunable disassortative property of this model,
which is supported by measurements in real technological networks
\cite{Internet}. The qualitative explanations of the correlations
and clustering spectrum can be found in \cite{BBV2}, and their
theoretical analysis appears in \cite{BP}. In sum, all the
simulation results for clustering and degree correlation, as
empirically observed, imply us that the increasing traffic may be
the driven force to shape the hierarchical and organizational
structure of real technological networks.

\subsection{Comparison with the BBV model}
One may notice that if the parameter $W$ is replaced by
$W=m\delta_{_A}$, then the mathematical framework of our
traffic-driven model is equivalent to that of BBV's \cite{BBV2},
though the specific weights' dynamics are quite different between
the two. By comparison with $\delta_{_{BBV}}$ (that is, the
fraction of weight which is locally ``induced" by the new edge
onto the neighboring others), the parameter $\delta_{_A}$ in our
model, with macroscopic perspectives, is the ratio of the total
weight increment on the expanding structure (W) to the number of
newly-established links (m) at each time step. It is an important
measure for the relative growing speed of traffic vs. topology,
and controls a series of the network scale-free properties. Thus,
there is an obvious difference between BBV model and ours: the
former is based on the local rearrangement of weights induced by
newly added links, while the latter is built upon the global
traffic growth and the redistribution of weights according to the
local nature of network. Noticeably, the weight dynamical
evolution of the BBV model is triggered only by newly added
vertices, hardly resulting in satisfying interpretations to
collaboration networks or the airport systems. For these two
cases, even if the size of networks keeps invariant, co-authored
papers will still come out and airports can become more crowded as
well. In contrast, this difficulty for practical explanations
naturally disappear in our model, due to its global weight
dynamics. Above all, the traffic-driven model here, without loss
of simplicity and practical senses, has generalized BBV's work and
narrowed its applicable scope to technological networks. Based on
it, more complicated variations of the microscopic rules may be
implemented to better mimic technological networks. Especially
worth remarking is that the present empirical studies on airport
networks and Internet indicate the nonlinear degree-strength
correlation $s\sim k^{\beta}$ with $\beta>1$. In Ref. \cite{BBV2},
Barrat \emph{et al.} proposed the heterogeneous coupling mechanism
to obtain this property. This is not difficult to introduce into
the present traffic-driven version. Moreover, we find that the
nontrivial weight-topology correlation can also emerge from the
accelerating growth of traffic weight \cite{BH} or from the
accelerating creation and reinforcement of internal edges
\cite{WWX, BH}.

\section{Model B}
\subsection{The Neighbor-Connected Model for Social Networks}
Social networks are a paradigm of the complexity of human
interactions, which have also attracted a great deal of attention
within the statistical physics community. The study of social
networks has been traditionally hindered by the small size of the
networks considered and the difficulties in the process of data
collection (usually from questionnaires or interviews). More
recently, however, the increasing availability of large databases
has allowed scientists to study a particular class of social
networks, the so-called collaboration networks. The co-authorship
network of scientists represents a prototype of complex evolving
networks, which can be defined in a clear way. Their large size
has allowed researchers to get a reliable statistical description
of their topological and weight properties, and hence reach
reliable conclusions of the network structure and dynamics. In
weighted social networks, the edge weight between a pair of nodes
can represent the tightness of their connection. The larger weight
indicates the more frequency of interaction; e.g. the number of
co-authored papers between two scientists, the frequency of
telephone or email contacts between two acquaintances, etc. In
addition, it is more probable that two vertices with a common
neighbor get connected than two vertices chosen at random.
Clearly, this property leads to a large average clustering
coefficient since it increases the number of connections between
the neighbors of a vertex. This is already observed in a model
proposed by Davidsen, Ebel and Bornhodt \cite{DEB}.

Our neighbor-connected model starts from an initial graph of $N_0$
vertices, fully connected by links with assigned weight $w_0=1$.
Its evolution then is simply based on the dynamics of
\emph{connecting nearest-neighbors}: At each time step, a new
vertex $n$ is added with one primary link and $m$ secondary links,
which connect with $m+1$ existing vertices (the initial network
configuration hence requires $N_0>m+1$). Actually, the newly-built
connections are not independent, but related with each other. The
primary link ($w_0=1$) first preferentially attaches to an old
node with large strength; i.e. a node $i$ is connected by the
primary link with probability
\begin{equation}
\Pi_{n\rightarrow i}=\frac{s_i}{\sum_ks_k}.
\end{equation}Then, the $m$ secondary connections
(assigned $w_0=1$ each) are preferentially built between the new
vertex and $m$ neighbors of node $i$, with the weight preferential
probability
\begin{equation}
\Pi_{n\rightarrow j}=\frac{w_{ij}}{s_i},
\end{equation}where $j\in\Gamma(i)$. After building the primary link $(n, i)$,
the creation of every secondary link $(n, j)$ is assumed to
introduce variations of network weights. For the sake of
simplicity, we limit ourselves to the case where the introduction
of a primary link on node $i$ will trigger only local
rearrangement of weights on the existing neighbors
$j\in\Gamma(i)$, according to the rule
\begin{equation}
w_{ij}\rightarrow w_{ij}+\delta.
\end{equation}In general, $\delta$ depends on the local
dynamics and can be a function of different parameters such as the
weight $w_{ij}$, the degree or the strength of $i$, etc. In the
following, we will simply focus on the case that $\delta=const$.
After the weights have been updated, the evolving process is
iterated by introducing a new vertex until the desired size of the
network is reached.

The above mechanisms have simple physical and realistic
interpretations. Once a fresh member joins a social community, he
will be introduced to his neighbors or take initiatives to
interact with them. Then, his social connections will soon be
built within his neighborhood. It is reasonable that the entering
of this new member can trigger the local variation of connections.
Take the SCN for example, a scientist joining a research group
will often collaborate with both the group director (primary link)
and the other group members (secondary links). The scientist's
affiliation can naturally boost the research productivity of the
group and also, enhance the collaborations of other members. The
above scenario may be a best interpretation of the origin of our
model parameter $\delta$. We use $\delta$ to control the effect of
the newly-introduced member on the weights of connections among
the local neighbors. In the proceeding section, we will see the
model also gives a wealth of scale-free behaviors depending on
this basic parameter.

\begin{figure}
\scalebox{0.80}[0.75]{\includegraphics{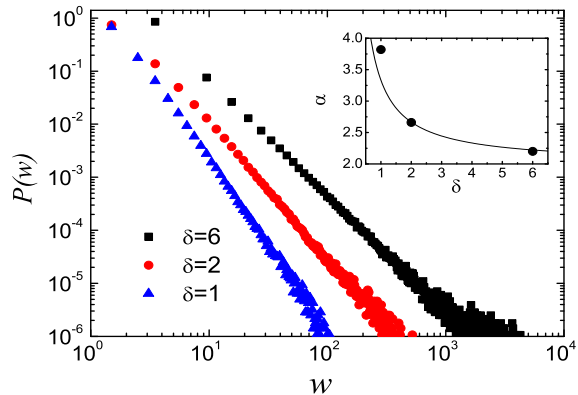}}
\caption{\label{fig:epsart} Weight probability distribution with
m=3. Data are consistent with a power-law behavior $P(w)\sim
w^{-\alpha_{_B}}$. In the inset we give the value of $\alpha_{_B}$
obtained by data fitting (filled circles), together with the
analytical form Eq. (22) (line). The data are averaged over 100
independent realizations of network size N=7000.}
\end{figure}

\subsection{Analytical and Numerical Results}
Along the analytical lines used in Section II, one can also
calculate the time evolution of the strength, weight and degree,
and hence calculated their scale-free distributions as follows:
\begin{equation}
\frac{dw_{ij}}{dt}=\frac{ms_i}{\sum_ks_k}\frac{w_{ij}}{s_i}\delta+
\frac{ms_j}{\sum_ks_k}\frac{w_{ij}}{s_j}\delta=2m\delta\frac{w_{ij}}{\sum_ks_k}.
\end{equation}By noticing $\sum_ks_k\approx2(m+1)t+2m\delta t=2(1+m+m\delta)t$, we have
\begin{equation}
\frac{dw_{ij}}{dt}=\frac{m\delta}{1+m+m\delta}\frac{w_{ij}}{t}=\theta\frac{w_{ij}}{t},
\end{equation}and so that $w_{ij}(t)=(t/t_{ij})^{\theta}$,
which indicates the power-law distribution of weight $P(w)\sim
w^{-\alpha_{_B}}$ with exponent
\begin{equation}
\alpha_{_B}=1+\frac{1}{\theta}=2+\frac{1+m}{m\delta}.
\end{equation}Further, the evolution equations for $s_i$ and $k_i$ are obtained
\begin{eqnarray}
\frac{ds_i}{dt}&=&\sum_j\frac{dw_{ij}}{dt}+\frac{s_i}{\sum_ks_k}+
\sum_{j\in\Gamma(i)}\frac{s_j}{\sum_ks_k}m\frac{w_{ij}}{s_j}
\nonumber\\
&=&2m\delta\frac{\sum_jw_{ij}}{\sum_ks_k}+(m+1)\frac{s_i}{\sum_ks_k}
\nonumber\\
&=&\frac{1+m+2m\delta}{2+2m+2m\delta}\frac{s_i}{t}=\lambda\frac{s_i}{t},\\
\frac{dk_i}{dt}&=&\frac{s_i}{\sum_ks_k}+\sum_{j\in\Gamma(i)}
\frac{s_j}{\sum_ks_k}m\frac{w_{ij}}{s_j} \nonumber\\
&=&\frac{m+1}{2+2m+2m\delta}\frac{s_i}{t}.
\end{eqnarray}Integrating with the initial conditions
$k_i(t=i)=s_i(t=i)=m+1$, we have
\begin{eqnarray}
s_i(t)&=&(m+1)\left(\zwfs{t}{i}\right)^{\lambda} \nonumber\\
k_i(t)&=&(m+1)\frac{s_i(t)+2m\delta}{1+m+2m\delta},
\end{eqnarray}which determine the scale-free distributions of both
strength and degree, with the same power-law exponent
\begin{equation}
\gamma_{_B}=1+\frac{1}{\lambda}=2+\frac{1+m}{1+m+2m\delta}.
\end{equation}

\begin{figure}
\scalebox{0.80}[0.75]{\includegraphics{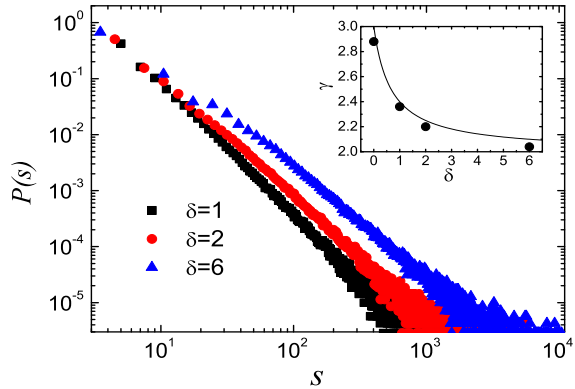}}
\caption{\label{fig:epsart} Strength probability distribution with
m=3. Data agree well with the power-law form $P(s)\sim
s^{-\gamma_{_B}}$. The inset reports $\gamma_{_B}$ from data
fitting (filled circles) averaged over 100 independent networks of
size N=7000, in comparison with the theoretical prediction Eq.
(26) (line).}
\end{figure}

\begin{figure}
\scalebox{0.85}[0.75]{\includegraphics{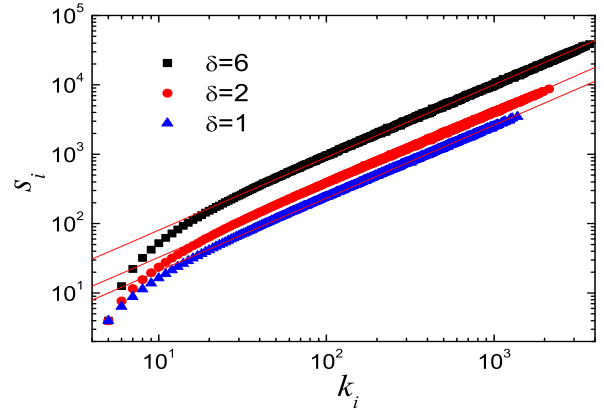}}
\caption{\label{fig:epsart} The average strength $s_i$ of vertices
with connectivity $k_i$ for different $\delta$ with m=3 (log-log
scale). Linear data fittings all give slope $1.00\pm0.02$,
confirming the prediction $\beta=1$.}
\end{figure}

We also performed numerical simulations of networks generated by
the model with various values of $\delta$ and minimum degree $m$.
As one can see, Fig. 6 and 7 recover the theoretical predictions
for scale-free distributions of weight and strength, and Fig. 8
validates the linear strength-degree correlation. It is worth
stressing that the empirical evidence in co-authorship networks
just indicates the linear correlation $s\sim k$ \cite{air2}.

\begin{figure}
\scalebox{0.80}[0.75]{\includegraphics{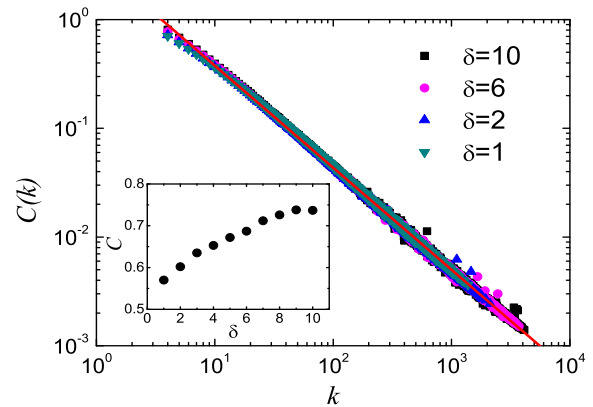}}
\caption{\label{fig:epsart} The scaling of $C(k)$ with $k$ for
various $\delta$ with m=3. Their fitted power-law exponents all
give 0.94 (dashed line). The data are averaged over 100
independent runs of network size N=7000. The inset shows the
average clustering coefficient $C$ depending on increasing
$\delta$.}
\end{figure}

\begin{figure}
\scalebox{0.80}[0.75]{\includegraphics{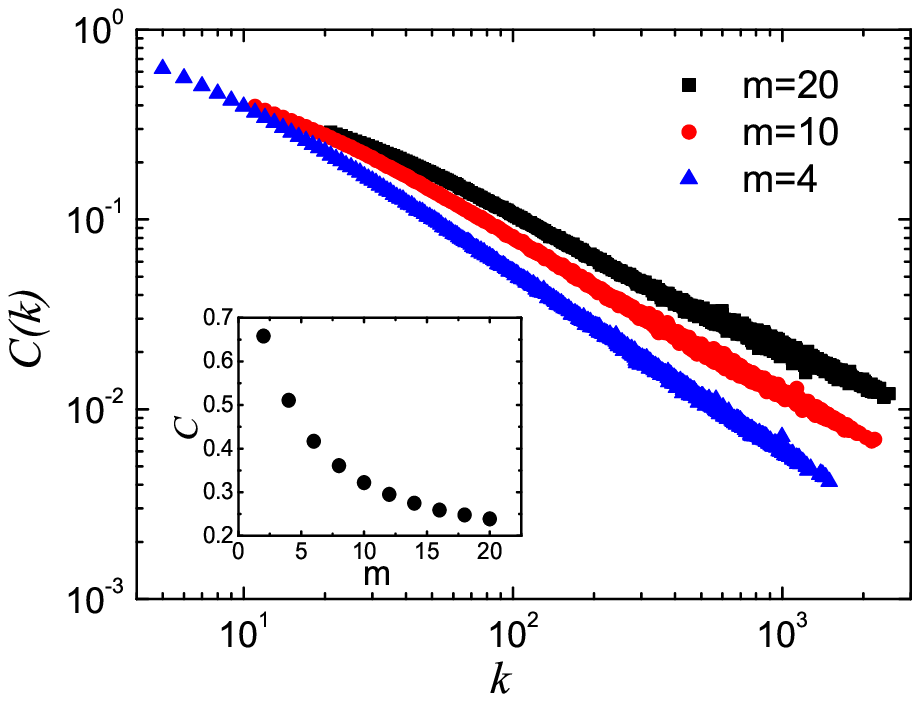}}
\caption{\label{fig:epsart} The scaling of $C(k)$ with $k$ for
various $m$ with $\delta=1$. The data are averaged over 100
independent runs of network size N=7000. The inset shows the
average clustering coefficient $C$ depending on increasing $m$.}
\end{figure}

\begin{figure}
\scalebox{0.80}[0.75]{\includegraphics{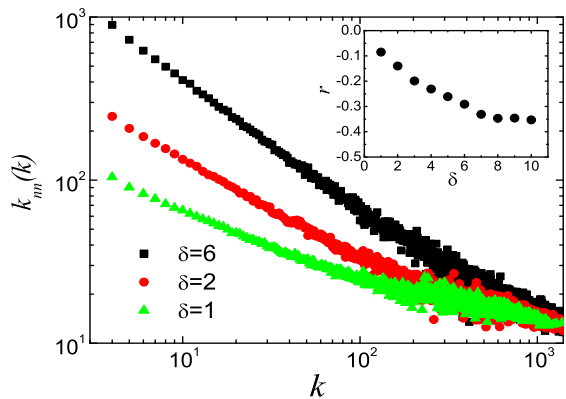}}
\caption{\label{fig:epsart} Average connectivity $k_{nn}(k)$ of
the nearest neighbors of a node depending on its connectivity $k$
for different $\delta$ with m=3. The data are averaged over 100
network realizations of N=7000. The assortativity coefficient $r$
versus $\delta$ is also reported in the inset.}
\end{figure}

\subsection{Discussions}
There are two important differences between the BBV model and our
neighbor-connected one, though the latter can be regarded as an
interesting variation of the former. First, in the BBV model the
evolution and distributions of such quantities as strength, weight
and degree are simply depending on the parameter $\delta_{_{BBV}}$
which controls the coupling between local topology and weights. In
our model, however, the evolution and distributions of those
quantities are controlled by two parameters ($\delta$ and $m$, as
reflected in theoretical part), which together determine the
effect of the new node on the local weights and topology. Though
we have fixed $m$ in most simulations, the role of parameter $m$
should not be ignored. As Fig. 9 reports, the curvature of $C(k)$
is not sensitive to the variations of $\delta$, but it
nontrivially depends on $m$ as shown in Fig. 10. Second, given the
same minimum degree, the average clustering coefficient $C$ of our
neighbor-connected model (see the inset of Fig. 9) can be much
larger than that of BBV's model or of our Model A, because the
secondary links in Model B considerably increase the density of
triangles within the system. Compared with the original model,
this larger clustering demonstrate an important advantage of Model
B in modelling the small-world property of real complex networks.
One question arises from the inset of Fig. 10: at first sight, it
may be surprising to see that $C$ greatly decreases when
increasing the secondary linking number $m$. Actually, larger $m$
in the model means larger minimum degree. When a new node arrives,
the number of triangles in the system will increase by $m$. But
for a smallest-degree node $i$, its clustering is
$c_i(m)=2N_\vartriangle(i)/m(m+1)$ where $N_\vartriangle(i)$
denotes the number of connected neighbors of node $i$. Therefore,
increasing $m$, we will decrease $c_i$ more greatly, resulting in
the decaying of $C$. Admittedly, there still exists a common point
which leads to the restriction of Model B to mimic and interpret
social networks. The degree correlations in both models are
negative (see the inset of Fig. 11), indicating their
disassortative mixing patten that is opposite in social networks.
Is it possible to find out a unified minimum model to characterize
both the assortative and disassortative networks? This question is
very challenging and appears among the leading ones in front of
network researchers \cite{top10}. Our recent studies may shed some
new light on this tackling problem \cite{WangHu}. Anyway, the
present neighbor-connected model (though disassortative) has
maintained the simplicity of the BBV model and appears as a more
specific version for weighted social networks, considering its
larger clustering and clearer hierarchical structure.

\section{Review and outlook}
In this paper, we have presented and studied two evolving models
for weighted complex networks. These two models intend to mimic
technological networks and social graphs respectively, and can be
regarded as a meaningful development of the original BBV model.
Though their specific evolution dynamics are different, all of
them are found to bear similar mathematical structures and hence
exhibit several common behaviors, e.g. the power-law distributions
and evolution of degree, weight and strength. In each case, we
also studied the nontrivial clustering coefficient $C$ and tunable
degree assortativity $r$ as well as their degree-dependent
correlations. In such context, we compared our generalized models
with the original one, and got several interesting conclusions,
which may provide us with a better understanding of the
hierarchical and organizational architecture of weighted networks.
For all the above reasons, we would like to classify these models
into a class, within which the BBV's is the first one and may be
the simplest one. It must be admitted that this class of models do
not take into account the internal connections during the network
evolution, which is yet beyond the scope of this paper. Still, we
hope that our generalized work can make this model family more
diversified and help bring it some new sights.


\begin{thebibliography}{ref1}
\bibitem{Internet} R. Pastor-Satorras and A. Vespignani, {\it Evolution
and Structure of the Internet: A Statistical Physics Approach}
(Cambridge University Press, Cambridge, England, 2004).

\bibitem{WWW} R. Albert, H. Jeong, and A.-L. Barab\'asi, Nature \textbf{401}, 130 (1999).

\bibitem{CN1} M.E.J. Newman, Phys. Rev. E \textbf{64}, 016132 (2001).

\bibitem{CN2} A.-L. Barab\'asi, H. Jeong, Z. N\'eda. E. Ravasz, A.
Schubert, and T. Vicsek, Physica (Amsterdam) \textbf{311A}, 590
(2002).

\bibitem{air1} R. Guimera, S. Mossa, A. Turtschi, and L.A.N Amearal, cond-mat/0312535.

\bibitem{air2} A. Barrat, M. Barth\'elemy, R. Pastor-Satorras, and
A. Vespignani, Proc. Natl. Acad. Sci. U.S.A. \textbf{101}, 3747
(2004).

\bibitem{BA} R. Albert and A.-L. Barab\'asi, Rev. Mod. Phys. \textbf{74}, 47 (2002).

\bibitem{top10} A virtual round tabel on ten leading questions for
network research can be found in the special issue on Applications
of Networks, edited by G. Caldarelli, A. Erzan and A. Vespignani
[Eur. Phys. J. B \textbf{38}, 143 (2004)].

\bibitem{ref1} W. Li and X. Cai, Phys. Rev. E \textbf{69}, 046106
(2004).

\bibitem{traffic-driven} K.-I. Goh, B. Kahng, and D. Kim,
cond-mat/0410078 (2004).

\bibitem{empirical} R. Pastor-Satorras, A. V\'azquez, and A.
Vespignani, Phys. Rev. Lett. \textbf{87}, 258701 (2001).

\bibitem{BBV} A. Barrat, M. Barth\'elemy, and A. Vespignani, Phys.
Rev. Lett. \textbf{92}, 228701 (2004).

\bibitem{WWX} W.-X. Wang, B.-H. Wang, B. Hu, G. Yan, and Q. Ou,
Phys. Rev. Lett. \textbf{94}, 188702 (2005).

\bibitem{mixing} M. E. J. Newman, Phys. Rev. E \textbf{67}, 026126
(2003).

\bibitem{hierarchy} E. Ravasz and A.-L Barab\'asi, Phys. Rev. E
\textbf{67}, 026112 (2003).

\bibitem{BBV2} A. Barrat, M. Barth\'elemy, and A. Vespignani, Phys. Rev. E \textbf{70}, 066149 (2004).

\bibitem{BP} A. Barrat and R. Pastor-Satorras, Phys. Rev. E
\textbf{71}, 036127 (2005).

\bibitem{BH} B. Hu \emph{et al.} (unpublished).

\bibitem{DEB} J. Davidsen, H. Ebel, and S. Bornhodt, Phys. Rev.
Lett. \textbf{88}, 128701 (2001).

\bibitem{WangHu} W.-X. Wang, B. Hu, B.-H. Wang, and G. Yan (unpublished).
\end{thebibliography}
\end{document}